\documentclass{tlp}

\usepackage[dvipsnames]{xcolor}
\usepackage{hyperref}
\usepackage{amsmath}
\usepackage{amssymb}
\usepackage{graphicx}
\usepackage{xspace}
\usepackage{verbatim}
\usepackage{fancyvrb}
\usepackage{inconsolata}
\usepackage{listings}
\usepackage{eurosym}

\newcommand{\naf}{\mathop{not}}
\newcommand{\clingo}{\texttt{clingo}}
\newcommand{\clasp}{\texttt{clasp}}
\newcommand{\gringo}{\texttt{gringo}}
\newcommand{\wasp}{\texttt{wasp}}
\newcommand{\dlv}{\texttt{DLV}}
\newcommand{\idlv}{\texttt{IDLV}}

\definecolor{color_5_services}{HTML}{0033cc}
\definecolor{color_10_services}{HTML}{FF7700}
\definecolor{color_15_services}{HTML}{cc0000}
\definecolor{color_20_services}{HTML}{28af45}

\newcommand{\subfigscale}{0.23}

\usepackage{subcaption}

\newcommand{\flexi}{{\sf FlexiPlace}\xspace}

\DefineVerbatimEnvironment{code}{Verbatim}
{gobble=4, fontfamily=zi4, fontsize=\footnotesize, frame=single, framesep=1mm, framerule=0.1pt, rulecolor=\color{gray}}

\DefineVerbatimEnvironment{codeNum}{Verbatim}
{gobble=4, fontfamily=zi4, numbers=left, numbersep=5pt, numberblanklines=false, firstnumber=last, tabsize=2, fontsize=\footnotesize, frame=single, framesep=1mm, framerule=0.1pt, rulecolor=\color{gray}, commandchars=\\\{\}}

\newtheorem{example}{Example}

\begin{document}

\title[Application Placement with Constraint Relaxation]{Application Placement with Constraint Relaxation}

\jnlPage{1}{8}
\jnlDoiYr{2021}
\doival{10.1017/xxxxx}


\lefttitle{Azzolini et al.}
\begin{authgrp}
\author{\sn{Azzolini} \gn{Damiano}}
\affiliation{University of Ferrara}
\email{damiano.azzolini@unife.it}
\author{\sn{Duca} \gn{Marco}}
\affiliation{University of Calabria}
\author{\sn{Forti} \gn{Stefano}}
\affiliation{University of Pisa}
\email{stefano.forti@unipi.it}
\author{\sn{Gallo} \gn{Francesco}}
\affiliation{University of Calabria}
\author{\sn{Ielo} \gn{Antonio}}
\affiliation{University of Calabria}
\email{antonio.ielo@unical.it}
\end{authgrp}

\history{\sub{xx xx xxxx;} \rev{xx xx xxxx;} \acc{xx xx xxxx}}

\maketitle              
\begin{abstract}  
Novel utility computing paradigms rely upon the deployment of multi-service applications to pervasive and highly distributed cloud-edge infrastructure resources. Deciding onto which computational nodes to place services in cloud-edge networks, as per their functional and non-functional constraints, can be formulated as a combinatorial optimisation problem. 
Most existing solutions in this space are not able to deal with \emph{unsatisfiable} problem instances, nor preferences, i.e. requirements that DevOps may agree to relax to obtain a solution. In this article, we exploit Answer Set Programming optimisation capabilities to tackle this problem. Experimental results in simulated settings show that our approach is effective on lifelike networks and applications.
\end{abstract}


\section{Introduction}
\label{sec:introduction}

In the last decade, cloud-edge computing paradigms (e.g., fog, edge, mist computing) have attracted increasing attention from both academic and industrial research communities (\cite{DBLP:journals/spe/Srirama24}). These paradigms extend the traditional cloud computing model by incorporating resources along a computing continuum, ultimately interconnecting Internet of Things (IoT) devices with cloud virtual machines through a hierarchy of intermediate layers spanning end-user devices, enriched infrastructure assets, and small-scale private data centres.
Overall, they aim at offering computing, storage and networking as utilities by leveraging a continuum of pervasive, heterogeneous resources that enable low-latency processing and context-aware service delivery, especially targeting latency-sensitive or bandwidth-intensive IoT applications, e.g., augmented reality, remote surgery or safety monitoring (\cite{vetriveeran2025resource,moreschini2022cloud}).

Deploying multi-service applications across cloud-edge resources comes with significant challenges due to the scale, heterogeneity and dynamic nature of resources, along with stringent Quality of Service (QoS) requirements of the applications to be deployed~(\cite{computers14030099}).
Such applications, composed of interacting services, must be placed to meet constraints such as latency, bandwidth, energy consumption, and locality of IoT devices.
Traditional placement strategies often approach this as a constraint satisfaction or optimisation problem~(\cite{MahmudApplicationManagement}).
In practice, certain combinations of constraints may be unsatisfiable, leading current methods to fail due to resource scarcity, too demanding application requirements, or both. Notably, the current literature on cloud-edge application placement does not address the case of requirements that should be \emph{preferably}, but not necessarily, satisfied according to given priorities.

For instance, a DevOps may request that a real-time video analytics service should be deployed to a node capable of reaching a surveillance camera with suitable latency and bandwidth, while also imposing strict limits on the carbon intensity of the chosen node.
If no node can satisfy all the constraints, existing approaches~(\cite{MahmudApplicationManagement}) would typically reject the deployment altogether.
This highlights the need for a placement strategy capable of reasoning over conflicting constraints and identifying which ones can be relaxed with \textit{minimal} impact, according to priorities defined by application DevOps.

In this context, we leverage Answer Set Programming~(\cite{brewka2011asp}) (ASP) for determining QoS-aware placements of multi-service applications within cloud-edge environments and propose a novel solution called \flexi.
The main novelty of our approach lies in its ability to manage unsatisfiable placement instances by selectively \emph{relaxing} (dropping) constraints based on a priority hierarchy established by the application DevOps (the ``domain experts''), allowing one to determine an eligible application deployment instead of failing.
We assess our tool on a set of benchmarks and show that it is effective on realistic-sized infrastructures and applications. 

The article is organised as follows. 
Section~\ref{sec:background} discusses the background, Section~\ref{sec:considered_problem} introduces the considered problem through a motivating scenario, which is encoded in ASP in Section~\ref{sec:encoding}.
Section~\ref{sec:experiments} presents the experimental evaluation, Section~\ref{sec:related} surveys related work, and Section~\ref{sec:conclusion} concludes the paper.

\section{Background}
\label{sec:background}

Answer Set Programming (ASP) is a popular declarative programming paradigm.
Its compact and expressive language makes it a powerful tool for handling knowledge-intensive combinatorial problems, both in industry and academia~(\cite{erdem2016applications,DBLP:journals/ki/FalknerFSTT18,azzolini2024continuous,DBLP:conf/lpnmr/BaumeisterHORRSW24,DBLP:conf/sc/GamblinCBS22}), also thanks to the availability of efficient reasoners~(\cite{gebser2019clingo,dlv}). 

\smallskip\noindent\textit{Syntax.}
A \emph{term} is either a \emph{variable}, a \emph{constant} or a \emph{function symbol}, where variables start with uppercase letters and constants start with lowercase letters or are numbers.
A function term is an expression of the form $f(t_1, \dots, t_n)$ where $f$ is its name and $t_i$ are terms. An \emph{atom} is an expression of the form $p(t_1,\ldots,t_n)$ where $p$ is a predicate of arity $n$ and $t_1,\ldots,t_n$ are terms; it is \emph{ground} if all its terms are constants.
A \emph{literal} is either an atom $a$ or its negation $\naf a$, where $\naf$ denotes the negation as failure.
A literal is said to be \emph{negative} if it is of the form $\naf a$, otherwise it is \textit{positive}.
For a literal $l$, $\overline{l}$ denotes the complement of $l$. More precisely, $\overline{l}=a$ if $l = \naf a$, otherwise $\overline{l}=\naf a$.
A \emph{normal rule} is an expression of the form $h \leftarrow b_1,\ldots,b_n$ where $h$ is an atom called \emph{head}. 
When $n\geq 0$, $b_1,\ldots,b_n$ is a conjunction of literals called \emph{body}.
A normal rule is said to be a \emph{constraint} if its head is omitted, while it is said to be a \emph{fact} if $n=0$.
A \emph{program} is a finite set of normal rules.
We will also use choice rules~(\cite{DBLP:conf/lpnmr/NiemelaSS99}). 
A \emph{choice element} is of the form $h:l_1,\ldots,l_k$, where $h$ is an atom, and $l_1,\ldots,l_k$ is a conjunction of literals.
A \emph{choice rule} is an expression of the form $\{e_1;\ldots;e_m\}\leftarrow b_1,\ldots,b_n$.
We also consider \textit{aggregate} atoms in the body of rules~(\cite{alviano2018aggregates}) of the form $\#sum\{\epsilon_0 ; \dots ; \epsilon_n\} \ > k$ where $k$ is called \textit{guard} and can be a constant or a variable and $\epsilon_0, \dots, \epsilon_n$ is such that each $\epsilon_i$ has the form $t_1, \dots, t_n : F$ and each $t_i$ is a term whose variables appear in the conjunction of literals $F$.

\smallskip\noindent\textit{Semantics.}
Given a program $P$ and $r \in P$, $ground(r)$ is the set of ground instantiations of $r$ obtained by replacing variables in $r$ with constants in $P$.
For aggregates, a variable is called \textit{local} if it appears only in the considered aggregate; \textit{global} otherwise.
The grounding of a rule with aggregates first requires replacing global variables and then replacing local variables appearing in aggregates with ground terms. 
We denote with $ground(P)$ the union of ground instantiations of rules in $P$.
An aggregate is true in an interpretation $I$ (i.e., a set of atoms) if the evaluation of the aggregate function under $I$ satisfies the guards.
We refer the reader to~\cite{DBLP:journals/tplp/CalimeriFGIKKLM20} for a more in-depth treatment of aggregates. 
Given a program $P$, an \textit{interpretation} $I$ is an \textit{answer set} (also called stable model) of $P$ iff $(i)$ $I$ is a model, i.e., for each rule $r \in ground(\Pi)$ either the head of $r$ is true w.r.t. $I$ or the body of $r$ is false w.r.t $I$; and $(ii)$ $I$ is a minimal model of its GL-reduct~(\cite{DBLP:journals/ngc/GelfondL91}).
If $P$ has no answer sets, it is called \textit{unsatisfiable}.

\smallskip\noindent\textit{Optimization.}
Weak constraints~(\cite{buccafurri2000weakconstraints}) are expressions of the form
$:\sim l_1, \dots, l_m.[w@p,t_1,\dots,t_n]$
where $l_1, \dots, l_m$ are literals, $w \in \mathbb{N}$ is the cost, $p \in \mathbb{N}$ is the priority, and $t_1, \dots, t_n$ is a tuple of terms.
Such rules associate each answer set with a \emph{cost} with a \emph{priority level}, which 
can be intuitively understood as an objective function, to be optimized in order of priority.
These enable one to tackle optimization problems in ASP~(\cite{carmineopt}).
An answer set with costs $c_0, c_1, \dots c_k$ has a ``lower cost'' than an answer set with costs $c_0', c_1', \dots, c_k'$ if there exist $i$ such that $c_i < c_i'$ and $c_j = c_j'$ for all $j < i$.
An answer set is \emph{optimal} if there does not exist an answer set with a lower cost.

\section{Motivating Scenario: the Application Placement Problem}
\label{sec:considered_problem}

In this section, we illustrate the considered problem by means of a simple, yet complete, motivating example adapted from the literature~(\cite{DBLP:conf/percom/Forti22}).
The depicted scenario epitomises a broader class of placement problems in which functional (e.g., hardware, IoT) and non-functional requirements related to sustainability (e.g., energy efficiency, carbon intensity), performance (e.g., latency, bandwidth), and reliability (e.g., availability, security) must be satisfied simultaneously, despite being often conflicting and constrained by limited resources. 
Typical deployments involve hundreds of services and nodes, leading to a combinatorial explosion of potential candidate solutions~(\cite{DBLP:journals/computing/SmolkaM22})
(e.g., mapping 30 services on 100 nodes yields up to $100^{30}=10^{60}$ candidates to check).

\begin{figure}
    \centering
    \includegraphics[width=0.6\linewidth]{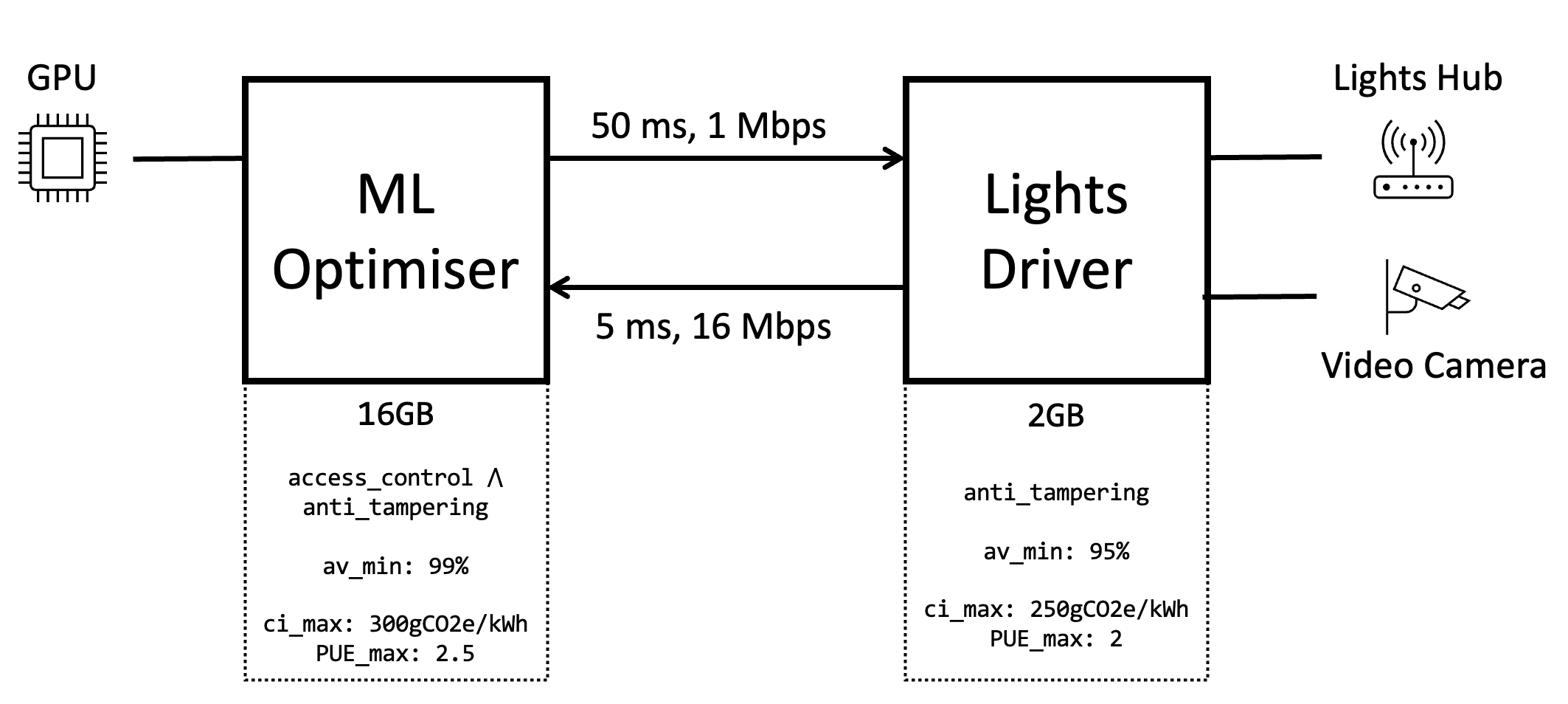}
    \caption{Example application.}
    \label{fig:app}
\end{figure}

The application in Figure~\ref{fig:app} manages street lighting using machine learning (ML) and includes two services: the \textsf{\small ML Optimiser}, which processes video streams to determine optimal lighting strategies, and the \textsf{\small Lights Driver}, which controls the street lights.
The \textsf{\small ML Optimiser} requires a GPU co-processor to train models that update the driver's control rules while the \textsf{\small Lights Driver} interfaces with both a lighting hub and a video camera, which monitors ambient conditions and streams footage to the optimiser.
Each service has functional and non-functional requirements.
For instance, the \textsf{\small ML Optimiser} requires 16 GB RAM, access control and anti-tampering mechanisms, minimum node availability (\texttt{av\_min}) of 99\%, carbon intensity of the node energy mix (\texttt{ci\_max}) below 300 gCO\textsubscript{2}-eq/kWh, and power usage effectiveness\footnote{PUE is a standard computing efficiency metric defined as the ratio of total energy consumption of an IT system to the energy used by computing equipment alone. A value of 1.0 indicates ideal efficiency.} (\texttt{PUE\_max}) under 2.5.
Additionally, communication constraints specify a maximum latency of 50 ms and a minimum bandwidth of 1 Mbps from the \textsf{\small ML Optimiser} to the \textsf{\small Lights Driver}, and 5 ms and 16 Mbps in the reverse direction.

\begin{figure}
    \centering
    \includegraphics[width=0.6\linewidth]{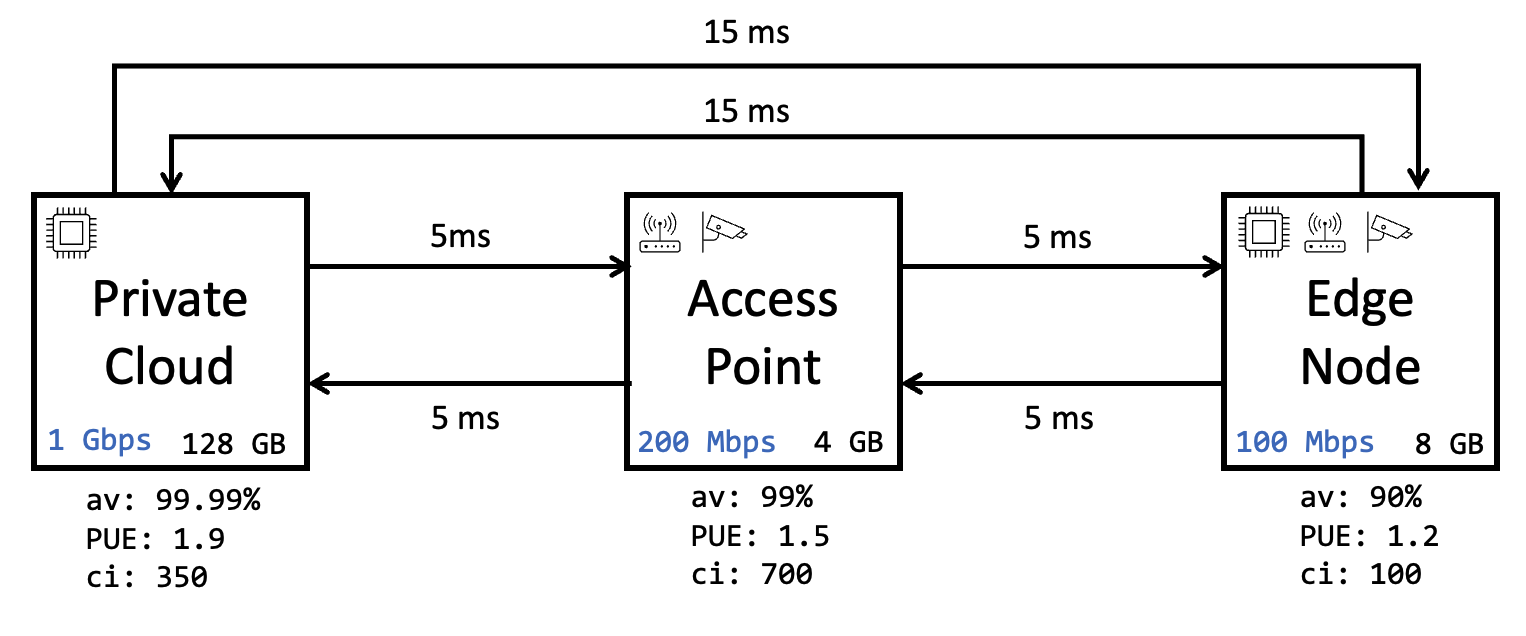}
    \caption{Example infrastructure.}
    \label{fig:infra}
\end{figure}

Figure~\ref{fig:infra} sketches the target infrastructure for the described application.
It consists of three interconnected computing nodes--\textsf{\small Private Cloud}, \textsf{\small Access Point}, and \textsf{\small Edge Node}, with a different availability of resources. For instance, the \textsf{\small Edge Node} is located closest to end devices, being directly connected to the lights hub and the video camera, and is equipped with 8 GB of RAM, a GPU, and 100 Mbps mobile connectivity. It has 90\% availability, a {PUE} of 1.2 and is powered up through an energy mix with a carbon intensity of 100 gCO\textsubscript{2}-eq/kWh. While the \textsf{\small Private Cloud} is assumed to feature all security mechanisms, the other two nodes are only equipped with an anti-tampering system to mitigate the damage in case of physical access to deployment resources.
Network communication latencies are symmetric and range from 5 ms between adjacent nodes to 15 ms between the \textsf{\small Private Cloud} and the \textsf{\small Edge Node}.

Determining an eligible placement of the application of Figure~\ref{fig:app} to the resources of Figure~\ref{fig:infra} requires mapping each service to a node that satisfies \textit{all} its functional and non-functional requirements, without exceeding the node's hardware and bandwidth capacity, which is an NP-hard problem~(\cite{QoSawaredeployment2017}).
In the described scenario, there is no solution placement that can satisfy all application requirements in the target infrastructure.
To address this, however, the DevOps team in charge of application deployment is willing to relax certain non-functional requirements (i.e., \textit{soft}) based on predefined priorities, where higher values indicate greater importance and less flexibility. Latency and bandwidth constraints have the highest priority (10) and are thus the least negotiable. Availability follows with priority 2, indicating moderate flexibility. Conversely, PUE and carbon intensity have the lowest priority (1). Security, hardware and IoT requirements remain non-negotiable (i.e. \textit{hard}) and must always be strictly enforced.

Hereinafter, we show how \flexi leverages ASP and accounts for
constraint prioritisation to relax a minimal set of
requirements
to compute feasible placements with the least impact on the original constraints.
That is, we aim at answering the following question: 
\textit{How can we determine a constraint-relaxed placement of a cloud-edge application that is both feasible and minimally deviates from our original deployment intent?}

\section{ASP Encoding}
\label{sec:encoding}

The core idea of our encoding is to match services to nodes in the infrastructure, discarding candidate solutions that do not satisfy \emph{application requirements} over infrastructure \emph{capabilities}.
As we will formalise in the following, we represent an infrastructure as a graph extended with \emph{attributes}.
Analogously, applications are represented as graphs and their requirements consist of simple arithmetic relationships between a (constant) threshold value for an attribute that a given service ``requires'' and the value of that specific attribute over the \emph{node where a service is placed}.
Rather than considering a fixed set of attributes for the nodes (and application requirements), our solution provides a way (by means of input facts) to \emph{define} the attributes of each node (and the corresponding requirements), that can be thus customized by users.

We first describe how to encode an input instance (infrastructure \& its capabilities, application \& its requirements) into a set of ASP facts, then how to model the application placement problem in ASP (``base encoding''), and finally how to address its relaxed version through ASP optimization, using the \textsc{clingo}~(\cite{gebser2019clingo}) input language. 

\subsection{Reification of Infrastructures \& Applications}

\smallskip\noindent\textit{Modelling the infrastructure.}
We model a target deployment infrastructure by means of predicates $node/1$ and $link/2$.
We use an atom $node(x)$ for each vertex $x$ in the network, and
an atom $link(x,y)$ for each edge that connects nodes $x$ and $y$.
Informally, both node attributes and link attributes are modelled as ``key-value pairs'', by means of predicates $node\_attr/3$ and $link\_attr/4$.
Atoms $node\_attr(x,k,v)$ mean that attribute $k$ has value $v$ on node $x$, and $link\_attr(x,y,k,v)$ mean that attribute $k$ on edge $(x,y)$ has value $v$.
This representation accommodates several kinds of infrastructure and application properties.

\begin{example}[Infrastructure]
The node \textsf{\small Private Cloud} of Figure~\ref{fig:infra} is encoded via the following set of facts:
\begin{code}
    node("prvt_cloud").
    node_attr("prvt_cloud","access_control",true). node_attr("prvt_cloud","anti_tampering",true). 
    node_attr("prvt_cloud","availability",9999).   node_attr("prvt_cloud","bandwidth_in",1000).
    node_attr("prvt_cloud","bandwidth_out",1000).  node_attr("prvt_cloud","carbon_intensity",350). 
    node_attr("prvt_cloud","gpu",true).            node_attr("prvt_cloud","pue",19). 
    node_attr("prvt_cloud","ram_gb",128). 
\end{code}
\noindent
Similarly, the link that connects such node to the  \textsf{\small Edge Node} is denoted by the fact:

\begin{code}
    link_attr("prvt_cloud", "edge_node", "latency", 15).
\end{code}
\end{example}

\smallskip\noindent\textit{Modelling the application.}
We use the predicates $service/1$ and $dependency/2$, with an analogous meaning to $node/1$ and $link/2$, to describe the application to be placed.
That is, $service(s)$ denotes that $s$ is a unique identifier for a service, and $dependency(s,t)$ that service $s$ ``depends on'' service $t$.
The counterpart of infrastructure attributes are service \emph{requirements}, which intuitively act as ``constraints''\footnote{In the rest of the section, we purposefully avoid using the word \emph{deployment constraints} as it is semantically overloaded with the notion of constraints in ASP.} to forbid deployments.
Intuitively, as services are deployed onto nodes and have pairwise dependencies, we can label each service with properties \emph{its matchee} must abide.
Thus, while infrastructure attributes refer to its vertices and edges, service requirements will refer to \emph{the node the service is deployed onto (during stable model search)}.  

A \emph{(simple) requirement} is a statement about infrastructure properties a node should possess in order to host a service. Here, we focus on requirement expressions that consist of comparisons between attributes' value and constants.
That is, expressions of the form $r \circ t$ where $r$ is an attribute, $\circ \in \{<, >, \le, \ge, =, \ne\}$ or $reserve(r,t)$. 
Intuitively, these respectively mean that the value (on a node) of an attribute $r$ should compare in a specific way against a threshold value $t$.
The requirement expression $reserve(r,t)$ expresses that $(i)$ attribute $r$ should be understood as consumable resources $(ii)$ a given service requires $t$ units of $r$ on the node to which it is deployed.
Application deployments often require that each service has access to a dedicated amount of computational resources (e.g., RAM or bandwidth).
Whenever multiple services are placed onto the same node or link asset, it must be ensured that the consumable resources are enough to host all services.

We encode requirement expressions by means of function terms such as $lt(r,v)$ (``less-than''), $gt(r,v)$ (``greater-than''), $eq(r,v)$ (``equals-to''), in addition to the aforementioned $reserve(r,v)$.
Given a requirement expression, we say that it holds on a given service using the atoms $hreq/2$ and $sreq/2$, which stand respectively for \textbf{h}ard \textbf{req}uirement and \textbf{s}oft \textbf{req}uirement.
The atom $hreq(s,e)$ denotes that the service $s$ can be deployed onto a node if and only if the node \emph{satisfies} the requirement $e$.
The atom $sreq(s,e)$ denotes that the service $s$ should be deployed onto a node that \emph{preferably} satisfies the requirement $e$. 
Similarly, the atoms $sreq((x,y),e)$ and $hreq((x,y),e)$ state that the requirement $e$ must be satisfied by the link between the nodes that host the services $x$ and $y$.

\begin{example}[Application requirements] 
The following encoding denotes the requirements of service {\sf\small ML Optimiser} of Figure~\ref{fig:app}:
\begin{code}
    hreq("ml_opt",eq("access_control",true)).  hreq("ml_opt",eq("anti_tampering",true)).
    sreq("ml_opt",gte("availability",99),2).   hreq("ml_opt",reserve("bandwidth_in",16)).
    hreq("ml_opt",reserve("bandwidth_out",1)). sreq("ml_opt",lte("carbon_intensity",300)).
    hreq("ml_opt",eq("gpu",true)).             sreq("ml_opt",lte("pue",25)). 
    hreq("ml_opt",reserve("ram_gb",16)).
\end{code}

Akin to infrastructure links, dependencies between services are denoted as:
\begin{code}
    sreq(("ml_opt", "lights_driver"), lte("latency", 50)). 
    sreq(("lights_driver", "ml_opt"), lte("latency", 5)).
\end{code}

\end{example}

\subsection{Encoding Application Deployment}
We now present the ``base'' encoding that solves the deployment problem.
It is based on a guess-and-check procedure, where choice rules guess a candidate assignment of services to infrastructure nodes and constraints prune assignments that violate requirements.
We denote such a program $\Pi_{deploy}$.
For now, \emph{we assume no distinction between hard requirements ($hreq/2$) and soft requirements ($sreq/2$)}. 
Given an application network $A$ and an infrastructure $R$, the answer sets of $\Pi_{deploy} \cup [R] \cup [A]$ can be mapped back to assignments that solve our problem.
A solution placement can be decoded by projecting answer sets onto the $deploy/2$ predicate.

\begin{figure}[t]
\begin{codeNum}
    resource(R) :- node_attr(_,R,_).
    \textcolor{OliveGreen}{% Deploy each service onto one node in the infrastructure.}
    \{ deploy(S,X): node(X) \} = 1 :- service(S).
    \textcolor{OliveGreen}{% hreq: hard requirements (can't be relaxed)}
    \textcolor{OliveGreen}{% sreq: soft requirements (can be relaxed)}
    req(S,E) :- hreq(S,E).
    req(S,E) :- sreq(S,E).
    \textcolor{OliveGreen}{% Cumulative usage of resource }
    shared_resource(R) :- req(_,reserve(R,_)).
    \textcolor{OliveGreen}{% Sum of all quantities Q of resource R reserved by services deployed in X}
    \textcolor{OliveGreen}{% must be below availability of R on Q}
    :- node_attr(X,R,T), shared_resource(R), #sum\{Q,S: deploy(S,X), req(S,reserve(R,Q))\} > T.
    :- req(S,reserve(R,Q)), deploy(S,X), node_attr(X,R,V), V < Q.
    \textcolor{OliveGreen}{% Enforcing requirement expressions (nodes).}
    :- req(S,eq(R,V)),  deploy(S,X), not node_attr(X,R,V).
    :- req(S,neq(R,V)), deploy(S,X), node_attr(X,R,V).
    :- req(S,lt(R,T)),  deploy(S,X), node_attr(X,R,V), V >= T.
    :- req(S,gt(R,T)),  deploy(S,X), node_attr(X,R,V), V <= T.
    :- req(S,gte(R,T)), deploy(S,X), node_attr(X,R,V), V < T.
    :- req(S,lte(R,T)), deploy(S,X), node_attr(X,R,V), V > T.
    \textcolor{OliveGreen}{% Enforcing requirement expressions (edges).}
    :- req((S1,S2),eq(R,V)),  deploy(S1,X), deploy(S2,Y),  not link_attr(X,Y,R,V).
    :- req((S1,S2),neq(R,V)), deploy(S1,X), deploy(S2,Y),  link_attr(X,Y,R,V).
    :- req((S1,S2),lt(R,T)),  deploy(S1,X), deploy(S2,Y),  link_attr(X,Y,R,V), V >= T.
    :- req((S1,S2),gt(R,T)),  deploy(S1,X), deploy(S2,Y),  link_attr(X,Y,R,V), V <= T.
    :- req((S1,S2),gte(R,T)), deploy(S1,X), deploy(S2,Y),  link_attr(X,Y,R,V), V < T.
    :- req((S1,S2),lte(R,T)), deploy(S1,X), deploy(S2,Y),  link_attr(X,Y,R,V), V > T.
\end{codeNum}
    \caption{\flexi main encoding.}
    \label{fig:fleximain}
\end{figure}

\subsection{Relaxing Soft Requirements on Application Deployment}
Distinguishing between hard and soft requirements naturally corresponds to an ASP optimization task.
The idea is to \emph{abduce over possible constraints} to remove atoms matching $sreq/2$ by means of choice rules; this disables the corresponding constraint in the logic program.
To do so,
$(i)$ we replace line 9 in the base encoding (Figure~\ref{fig:fleximain}) with the rules in Figure~\ref{fig:flexiopt} and 
$(ii)$ if we wish to weight (e.g., assign a preference score) to soft requirements to remove, we introduce atoms $violation\_cost(S,E,(C,L))$ to denote that we will pay a cost of $C$ at level $L$ if we renounce the soft requirement $sreq(S,E)$.

\begin{figure}[t]
\begin{codeNum}
    \{ req(S,E) \} :- sreq(S,E,_). 
    lift(S,E) :- sreq(S,E), not req(S,E).
    :~ violation_cost(S,E,(C,L)), lift(S,E). [C@L,S,E]
\end{codeNum}
    \caption{Additions to the main encoding of Figure~\ref{fig:fleximain} to address the relaxed problem.}
    \label{fig:flexiopt}
\end{figure}

Optimal answer sets correspond to optimal solutions of the deployment problem, where atoms $\mathit{lift}(S,E)$ denote that we renounce the requirement $E$ on the deployment of service $S$.
If this logic program is unsatisfiable, it means that even removing all $sreq$s, this \emph{would not be sufficient to ensure existence of a deployment}.
That is, (at least one of) the reason(s) for the inconsistency lies in $hreq$s alone, and further analysis would be required.
The logic program is satisfiable if there exists an assignment that perfectly fits all the requirements.
As we are interested in detecting soft requirements to relax rather than finding deployments, we consider projection of answer sets on the $\mathit{lift}/2$ predicate.
From the ASP modelling point of view, it would have been equivalent (in terms of optimal solutions) to directly express soft requirements as weak constraints.
However, our design choices have several practical advantages: 
$(i)$ the $\mathit{lift/2}$ predicate enables to easily inspect answer sets and retrieve \emph{which} requirements have been relaxed to achieve the solution,
$(ii)$ it is possible to control by means of facts (i.e., those matching $sreq/2$ and $hreq/2$) which requirements are mandatory and which ones can be relaxed, and 
$(iii)$ atoms $\mathit{lift/2}$ could be naturally used as \emph{objective atoms} to compute minimal unsatisfiable subprograms for explainability purposes~(\cite{DBLP:journals/ai/AlvianoDFPR23}).

\begin{example}
Consider again the motivating scenario of Section~\ref{sec:considered_problem}.
As discussed above, there is no eligible placement that meets all the requirements for the application of Figure~\ref{fig:app} to the infrastructure of Figure~\ref{fig:infra}.
For instance, the only node that can support the execution of the {\sf\small ML Optimiser} (i.e., the {\sf\small Private Cloud} node) features a carbon intensity of 350 gCO$_2$-eq which exceeds the required 300 gCO$_2$-eq.
Note that our model relies on facts like 
\begin{code}
    violation_cost(("ml_optimiser","lights_driver"),lte("latency",50),(10,1)).
\end{code}
to set the cost (i.e. priority) for relaxing soft constraints.
Running the encoding of Figures~\ref{fig:fleximain} and~\ref{fig:flexiopt} over the input denoting our motivating scenario returns an eligible placement that suggests deploying {\sf\small ML Optimiser} to {\sf\small Private Cloud} and {\sf\small Lights Driver} to {\sf\small Access Point}, and is obtained by relaxing constraints on carbon intensity for both services.
Such a solution is optimal, as it only relaxes two of the lowest-priority constraints, as indicated by the DevOps team in charge of managing the application.    
\end{example}

\section{Experiments}

\label{sec:experiments}

\begin{table}[t]
    \caption{Considered node properties. The table does not contain \emph{latency} since it is a link property. The ``Shared'' column denotes whether the property is considered \emph{shared among all services} hosted on the considered nodes. 
    The ``Constr'' column reports the constraint instantiated for the attribute during the generation process, for each service. 
    }
    \label{tab:network_properties}
    \centering
    \begin{tabular}{c|c|c|c || c|c|c|c }
        Attribute & Type  & Shared & Constr & Attribute & Type  & Shared & Constr\\
        \hline
        Access Control & Bool  & No & Equals  & CPU & Int  & Yes & Reserve \\
        Anti-tampering & Bool  & No & Equals  & Encryption & Bool  & No & Equals\\
        Availability & Int  & No & At Least & GPU & Bool  & No & Equals\\
        Bandwidth (In) & Int  & Yes & Reserve & Latency & Int  & No & At Most \\
        Bandwidth (Out) & Int  & Yes & Reserve &  PUE & Int  & No & At Most \\
        Carbon Intensity & Int  & No & At Most & RAM & Int  & Yes & Reserve\\
        Cost & Int  & No & At Most & Storage & Int  & Yes & Reserve\\
         \hline
    \end{tabular}
\end{table}

We perform a set of experiments to assess the effectiveness of our approach in computing cost-optimal relaxed deployments.
We first analyse the behaviour of available ASP systems on our encoding. 
Then, we conducted a more in-depth analysis to investigate the trade-off between model-guided~(\cite{bbalgo}) and core-guided~(\cite{uscalgo}) optimization algorithms for our application scenario. 

The experiments were run on a server with Intel(R) Xeon(R) CPU E7-8880 v4 @ 2.20GHz CPU, equipped with 500GB RAM, with a timeout of 180 seconds using GNU Parallel and executing at most 16 jobs in parallel. All code and data are available at \url{https://github.com/ainnoot/asp-app-placement}. A timeout in our setting refers to not being able to find an \textit{optimal model} or proving unsatisfiability within 180 seconds.

\paragraph{Data.}
We provide an instance generator for the problem, following standard practice in literature~(\cite{DBLP:journals/logcom/FortiBB22,004,DBLP:journals/eswa/Ghobaei-AraniS22}). 
We focus on the infrastructure attributes in Table~\ref{tab:network_properties} and define a set of realistic templates for infrastructure nodes and for application services. 
Infrastructure graphs and application graphs are obtained by sampling from Barab\'asi-Albert~(\cite{barabasi1999emergence}) and Erd\H os-Renyi~(\cite{erdosrenyi}) topologies, respectively.
On one hand, the Barab\'asi-Albert model captures the scale-free property of real-world ICT networks, where node degree distribution usually follows a power-law~(\cite{DBLP:books/ox/Newman10}).
This reflects the heterogeneity and hierarchy typical of cloud-edge infrastructures, where a small number of nodes act as high-bandwidth hubs and others as resource-constrained peripheral nodes.
On the other hand, the Erd\H{o}s-Renyi model neutrally approximates application topologies, where dependencies between services are established with uniform probability~(\cite{DBLP:books/ox/Newman10}).
This reflects the loosely coupled and stochastic nature of microservice-based applications, where interactions do not follow any hierarchical patterns~(\cite{DBLP:journals/access/VelepuchaF23,DBLP:journals/jss/SoldaniTH18}). 
Each node is assigned (uniformly at random) a ``configuration'' from a finite set.
In our case, the only link attribute is latency, which we model as a random integer between 10 and 50.
Nodes that are not directly connected by an edge are assigned an edge with a latency equal to the sum of the latencies in the shortest path between the two nodes.
As an example, if a generated graph contains the edges $(a,b)$ and $(b,c)$, we introduce the edge $(a,c)$ with latency $\ell(a,b) + \ell(b,c)$, where $\ell(x,y)$ is the latency on edge $(x,y)$. 
Following these procedures, we generated 10 infrastructures of size $\{50, 100, 150, \dots, 500\}$, and 6 applications of size $\{5, 10, 15, \dots, 30\}$.
For each combination, we generated 100 input pairs (i.e., application and infrastructure).
This yields a total of $10 \cdot 6 \cdot 100 = 6 \cdot 10^3$ instances.
We denote by $I(n,k)$ the set of instances considering the deployments of an application of size $k$ over infrastructures of size $n$.
We also use the notation $I(\cdot,\{k_0,k_1,\dots\})$ to denote ``all problem instances that deal with applications of size $k_0$, $k_1,$, $\dots$'' and $I(\{n_0,n_1,\dots\},\cdot)$ to denote ``all problem instances that deal with infrastructures of size $n_0$, $n_1$, $\dots$''.
Lastly, each of the considered applications consists of constraints over \emph{all} infrastructure properties defined in Table~\ref{tab:network_properties}.
Note that we use a weight of $1$ for all constraints that can be relaxed, with a single priority level, but the encoding we provide is more general.

\subsection{Solver Selection}
\label{sec:exp1}
We consider four ASP systems, obtained by pairing up the ASP solvers \clasp{}~(\cite{clasp}) and \wasp{}~(\cite{wasp}) with the ASP grounders I-DLV~(\cite{idlv}) and \gringo{}~(\cite{gringo}).
Note that the \gringo{}+clasp and \idlv{}+\wasp{} combinations are, essentially, the combinations adopted in the \clingo{}~(\cite{gebser2019clingo}) and \dlv~(\cite{dlv}) solvers.
We refer to each system as \gringo{}+\wasp{}, \gringo{}+\clasp{}, \idlv{}+\wasp{}, \idlv{}+\clasp{}.
We execute the systems using both a \emph{model-guided} (BB)~(\cite{bbalgo}) and a \emph{unsatisfiable core-guided} (USC)~(\cite{uscalgo}) algorithm, which typically yield complementary performances~(\cite{carmineopt}). 
In brief, model-guided algorithms attempt to iteratively improve lower bound solutions, \textit{à la} branch \& bound, while unsatisfiable core-guided algorithms try to treat all weak constraints as standard (strong) constraints, using unsatisfiable cores found within the optimization routine to shrink the search space.
We consider the \texttt{bb} and \texttt{oll} algorithms for \clasp{}, and the \texttt{basic} and \texttt{one} algorithms for \wasp{}, which are the default for the model-guided and core-guided algorithms in these solvers, respectively.
This yields a total of 8 configurations 
that we run over the 10\% of the total instances, selecting 60 instances for each network size, for a total of 600 instances.

\begin{figure}
\centering
\includegraphics{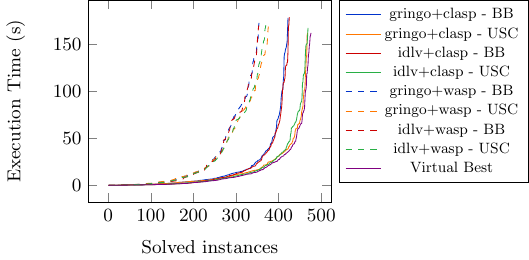}
\caption{Solvers comparison.}
\label{fig:cactus_solver_comparison_basic_one}
\end{figure}

\begin{figure}[t]
\centering
\begin{minipage}[c]{0.48\linewidth}
\centering
\includegraphics{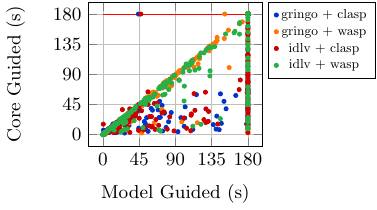}
\end{minipage}
\hfill
\begin{minipage}[c]{0.48\linewidth}
\centering
\includegraphics{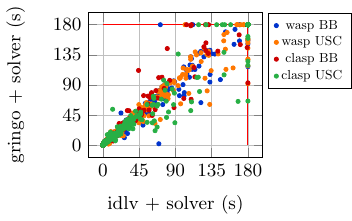}
\end{minipage}
\caption{Left: a point $(x,y)$ denotes that a given problem instance is solved in $x$ seconds using the model-guided and in $y$ seconds using the core-guided algorithm.
Right: a point $(x,y)$ denotes that a given problem instance is solved in $x$ seconds using the \idlv{} and in $y$ seconds using \gringo{} grounder.}
\label{fig:scatter_solvers_configurations}
\end{figure}

Figure~\ref{fig:cactus_solver_comparison_basic_one} shows that, overall, \clasp{}-based configurations outperform all \wasp{}-based configurations in terms of execution times.
In particular, the core-guided configuration of the \gringo{}+\clasp{} system (i.e., \clingo{} with default parameters) essentially overlaps with the \emph{virtual best solver}.
Recall that the virtual best solver is a fictitious system that is assumed to perform (instance-wise) as the best among the available solvers.
The scatter plots provide an instance-wise comparison of the systems.
We can observe in 
Figure~\ref{fig:scatter_solvers_configurations} (left)
that the default core-guided algorithm outperforms the default model-guided algorithm across all systems.
Moreover, 
Figure~\ref{fig:scatter_solvers_configurations} (right)
confirms that the grounder plays a less important role in our problem, with points distributed along the bisector.

\begin{figure}[t]
\centering
\begin{minipage}[c]{0.48\linewidth}
\centering
\includegraphics{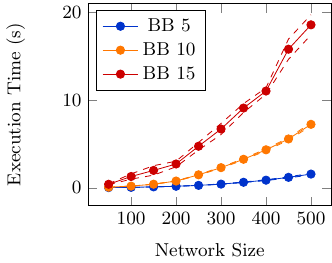}
\end{minipage}
\hfill
\begin{minipage}[c]{0.48\linewidth}
\centering
\includegraphics{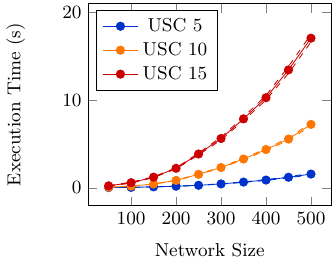}
\end{minipage}
\caption{Mean (solid) and standard error (dashed) on execution time to find the first optimal solution over $I(\cdot, \{5,10,15\})$ using BB (left) and USC (right) algorithms.}
\label{fig:easyinstances}
\end{figure}

\subsection{Assessment of \flexi}
From the previous set of experiments the \gringo{}+\clasp{} system (that is, \clingo{}) obtained the best overall performance among the tested configurations.
Thus, we focus on that system and perform a more in-depth assessment of our approach on \textit{all} the generated instances ($6 \cdot 10^3$).
We start by discussing the easy instances and then continue with an in-depth analysis of the results in the harder instances. Memory-wise, in all settings, we do not report significant memory usage.

\smallskip\noindent\textit{Easy instances.}
Applications of size up to 15 yield solvable instances for both optimization techniques.
Figure~\ref{fig:easyinstances} reports average runtime (up to the first optimal solution) over instances $I(\cdot,\{5,10,15\})$, where we can observe indeed an exponential-like effect of the application size w.r.t. infrastructure size on the overall runtime regardless of the solving algorithm.
Here, BB and USC have similar performances.
Overall, this is already sufficient to show applicability of our technique on non-trivial deployments, over realistic-sized infrastructures.
We continue our analysis, focusing on deployments with 20, 25 and 30 applications, being these more challenging. 

\begin{figure}
    \centering
\begin{minipage}{\textwidth}
\centering
\begin{minipage}{0.49\textwidth}
    \centering
\resizebox{0.83\textwidth}{!}{
    \begin{tabular}{c || c|c || c|c || c|c }
    \multicolumn{1}{c}{} & \multicolumn{6}{c}{\# services} \\
        \multicolumn{1}{c}{} & \multicolumn{2}{c}{20} & \multicolumn{2}{c}{25} & \multicolumn{2}{c}{30} \\
        size & BB & USC & BB & USC & BB & USC \\
        \hline
50 & 28 & 3 & 35 & 38 & 28 & 37 \\
100 & 39 & 6 & 82 & 50 & 99 & 94 \\ 
150 & 32 & 4 & 83 & 49 & 98 & 94 \\
200 & 18 & 0 & 68 & 28 & 96 & 89 \\
250 & 24 & 1 & 75 & 28 & 99 & 95 \\
300 & 21 & 3 & 62 & 37 & 96 & 89 \\
350 & 22 & 1 & 68 & 40 & 96 & 91 \\
400 & 19 & 5 & 76 & 46 & 100 & 95 \\
450 & 25 & 14 & 64 & 44 & 93 & 89 \\
500 & 21 & 6 & 65 & 43 & 93 & 83 \\
         \hline
    \end{tabular}
    }
\caption{Number of timeouts per application size and network size.}
\label{tab:table_unsolved}
\end{minipage}
\hfill
\begin{minipage}{0.49\textwidth}
\centering
\includegraphics{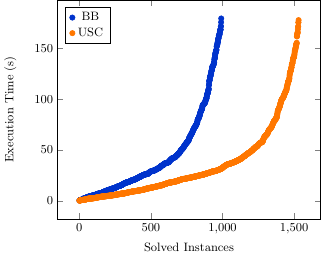}
\caption{Cumulative runtime of BB and USC algorithms over all instances.}
\label{fig:cactus}
\end{minipage}
\end{minipage}
\label{fig:table_and_cactus}
\end{figure}

\begin{figure}[t]
\centering
\begin{subfigure}{0.3\textwidth}
\centering
\includegraphics{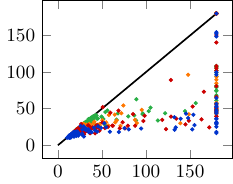}
\caption{20 services.}
\end{subfigure}
\hfill 
\begin{subfigure}{0.3\textwidth}
\centering
\includegraphics{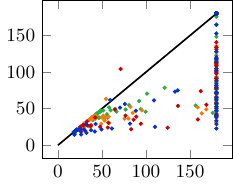}
\caption{25 services.}
\end{subfigure}
\hfill 
\begin{subfigure}{0.3\textwidth}
\centering
\includegraphics{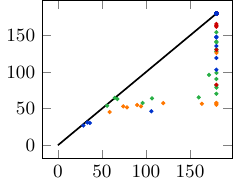}
\caption{30 services.}
\end{subfigure}
\caption{Execution time over instances in 
(a) $I(\{\textcolor{color_5_services}{350}, \textcolor{color_10_services}{400}, \textcolor{color_15_services}{450}, \textcolor{color_20_services}{500}\}, 20)$,
(b) $I(\{\textcolor{color_5_services}{350}, \textcolor{color_10_services}{400}, \textcolor{color_15_services}{450}, \textcolor{color_20_services}{500}\}, 25)$, and 
(c) $I(\{\textcolor{color_5_services}{350}, \textcolor{color_10_services}{400}, \textcolor{color_15_services}{450}, \textcolor{color_20_services}{500}\}, 30)$.
A point $(x,y)$ denotes that the USC algorithm solves an instance in $y$ seconds, while BB solves it in $x$ seconds.
Colors denote the size of the underlying network on top of which the applications are deployed.}
\label{fig:scatter_plots}
\end{figure}

\smallskip\noindent\textit{Harder instances.}
Instances $I(\cdot,\{20,25,30\})$ yield more interesting behaviour and deserve further analysis.
First, we observe in Table~\ref{tab:table_unsolved} that several instances hit the time-limit, with both optimization algorithms. 
Figure~\ref{fig:cactus} reports the overall performance of the two optimization algorithms over these instances, in terms of a cactus plot.
Overall, we can observe that these instances are better suited to be solved with USC techniques, as it is able to solve many more instances to optimality.
Instance-wise, the scatter plot in Figure~\ref{fig:scatter_plots} confirms the result, usually with USC outperforming BB.
However, USC also accrues more and more time-limits as the application size increases. 

\begin{figure}
\centering
\begin{subfigure}{\subfigscale\textwidth}
\includegraphics{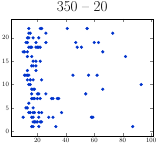}
\end{subfigure}
\hfill
\begin{subfigure}{\subfigscale\textwidth}
\includegraphics{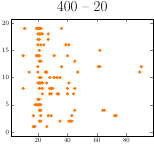}
\end{subfigure}
\hfill
\begin{subfigure}{\subfigscale\textwidth}
\includegraphics{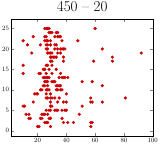}
\end{subfigure}
\hfill
\begin{subfigure}{\subfigscale\textwidth}
\includegraphics{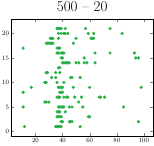}
\end{subfigure}
\begin{subfigure}{\subfigscale\textwidth}
\includegraphics{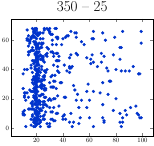}
\end{subfigure}
\hfill
\begin{subfigure}{\subfigscale\textwidth}
\includegraphics{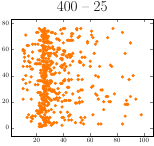}
\end{subfigure}
\hfill
\begin{subfigure}{\subfigscale\textwidth}
\includegraphics{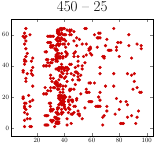}
\end{subfigure}
\hfill
\begin{subfigure}{\subfigscale\textwidth}
\includegraphics{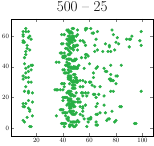}
\end{subfigure}


\begin{subfigure}{\subfigscale\textwidth}
\includegraphics{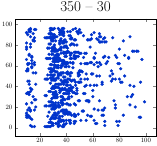}
\end{subfigure}
\hfill
\begin{subfigure}{\subfigscale\textwidth}
\includegraphics{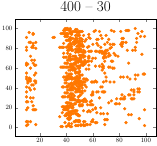}
\end{subfigure}
\hfill
\begin{subfigure}{\subfigscale\textwidth}
\includegraphics{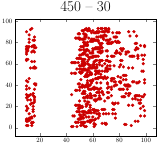}
\end{subfigure}
\hfill
\begin{subfigure}{\subfigscale\textwidth}
\includegraphics{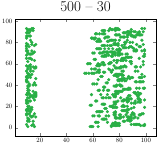}
\end{subfigure}
\caption{Temporal distribution of answer sets using the BB algorithm. Each cell reports, in a dotted chart, data about $I(n,k)$ (with label $n$---$k$). A point $(x,y)$ denotes that invoking the ASP solver on the $y$-th instance of $I(n,k)$ yields a model at time $x$.}
\label{fig:dotchart-scatters}

\end{figure}

\smallskip\noindent\textit{Temporal Distribution of Sub-optimal Answer Sets in BB.}
In practical scenarios, obtaining sub-optimal solutions in a fast way might still be useful.
Thus, one might be interested in investigating whether sub-optimal solutions are obtained at all, whenever optimal solutions are unavailable.
Figure~\ref{fig:dotchart-scatters} provides a plot on how non-optimal answer sets are found and distributed within allowed runtime when using the model-guided BB algorithm.
We observe that overall \emph{some} solutions (for all instances) are found within the first minute, then answer sets become more sparse, up to timeout---that typically occurs whenever the solver ``hits an optimal model'', but is not able yet to certify it as optimal (e.g., proving non-existence of a model with lesser cost).

Overall, USC and BB performance are comparable over 5-10-15 instances, regardless of network size, while USC is generally preferable for ``harder'' instances in the 20-25-30 range. However, BB is a way to obtain sub-optimal solutions quickly, whereas USC would time-out. We remark that, in a real-world setting, it would be totally feasible to run both approaches in parallel, so as to pick the first (optimal) solution found by either approach.

\section{Related Work}
\label{sec:related}

As aforementioned, the problem of deciding how to place application services to cloud-edge nodes in a QoS- and context-aware manner has been thoroughly studied. Here, we focus on the most closely related work, and we refer the readers to recent surveys by~\cite{computers14030099},~\cite{DBLP:journals/csur/PallewattaKB23}, and~\cite{DBLP:journals/csur/Ait-SalahtDL20} for further details.

Many solutions exist that rely on different techniques to determine application placements that meet functional and non-functional requirements.
Among these, most of the approaches relied on 
informed (heuristic) search~(\cite{QoSawaredeployment2017,004}), 
mathematical programming~(\cite{DBLP:journals/soca/SkarlatNSBL17,DBLP:journals/jpdc/MahmudSRB20}), 
bio-inspired meta-heuristics~(\cite{DBLP:journals/eswa/Ghobaei-AraniS22}), 
and deep learning solutions~(\cite{DBLP:journals/tmc/GoudarziPB23}).
Different solutions aims at optimising one (or more) aspect(s) of application placements, e.g., operational costs, latency, energy consumption, and resource usage.
Logic programming solutions, mainly written in Prolog, have recently been proposed to tackle the application placement problem, with a focus on aspects such as data locality~(\cite{DBLP:conf/summersoc/Massa0B22}), 
security and trust requirements~(\cite{secfog2019}), 
environmental impact~(\cite{DBLP:conf/padl/FortiB22}), or high-level network intent satisfaction~(\cite{DBLP:conf/icin/MassaFPDB24}). Notably, by classifying intent properties as either hard or soft, the latter approach supports recommending changes to original intents aimed at resolving emerging conflicts.

\cite{DBLP:journals/logcom/FortiBB22} relied on continuous reasoning mechanisms to enable incremental updates of solution placements in response to changes in application requirements or infrastructure capabilities, rather than recomputing solutions from scratch.
On a similar, yet complementary line,~\cite{azzolini2024continuous} proposed a solution combining ASP optimisation and Prolog-based continuous reasoning to distribute container images in cloud-edge settings.
ASP is also adopted by~\cite{dcop}, where the authors addressed the distributed constraint optimization problem.
The constraint-based approach of~\cite{DBLP:conf/lopstr/AmadiniGSVBFGPPZ24} complements our solution by addressing sustainable cloud-edge application placement via adaptive service flavour and topology selection under cost and carbon constraints. 

Similarly to other declarative programming efforts, \flexi allows modelling constraints including, e.g., hardware resources, availability, bandwidth, security policies, inter-component latency PUE, and carbon intensity.
Differently from all previous work, it features the possibility of automatically relaxing 
lower-priority constraints when no feasible deployment can be determined.
This flexibility goes beyond the state of the art, by implementing a graceful degradation of determined solution placements while guaranteeing critical constraints are met.
To the best of our knowledge, \flexi is the first approach integrating
logic-based placement with priority-based constraint relaxation.

\section{Concluding remarks}
\label{sec:conclusion}

We proposed a declarative approach based on ASP for placing multi-service applications in cloud-edge environments, and its open-source prototype \flexi. Our solution addresses satisfiable instances, allowing their declarative specification through a customisable and extensible taxonomy. Besides, it extends the state-of-the-art by solving unsatisfiable application placement instances through the selective relaxation of lower-priority constraints, according to priorities set by DevOps. Experimental results confirm the feasibility of the approach on realistic infrastructures (up to 500 nodes) and applications (up to 30 services).
As future work, we plan to support more expressive requirements and integrate explanations for unsatisfiable placement instances by relying on the notion of \emph{minimal unsatisfiable subprogram}~(\cite{DBLP:journals/ai/AlvianoDFPR23}).

\section*{Acknowledgments}
This work has been partly supported by project ``SEcurity and RIghts In the CyberSpace - SERICS'' (PE00000014 - CUP H73C2200089001) under the National Recovery and Resilience Plan (NRRP) funded by the European Union - NextGenerationEU. DA is a member of the Gruppo Nazionale Calcolo Scientifico -- Istituto Nazionale di Alta Matematica (GNCS-INdAM).

\bibliographystyle{acmtrans}
\bibliography{bibl}

\end{document}